\documentclass[conference]{IEEEtran}
\usepackage{amssymb}
\usepackage{graphicx}
\usepackage[tbtags]{amsmath}
\usepackage{amsfonts}
\usepackage{amssymb}

\newtheorem{thm}{Theorem}

\newtheorem{lemma}[thm]{Lemma}
\newtheorem{definition}[thm]{Definition}
\newtheorem{proposition}[thm]{Proposition}

\newtheorem{remark}{Remark}

\newcommand{\renyi}{R\'{e}nyi }
\newcommand{\csiszar}{Csisz\'{a}r }

\begin{document}
\title {Further Results on Geometric Properties of a Family of Relative Entropies}
\author{
\IEEEauthorblockN{Ashok Kumar M.}
\IEEEauthorblockA{Department of ECE \\
Indian Institute of Science \\
Bangalore, Karnataka 560012, India\\
Email: ashokm@ece.iisc.ernet.in}
\and
\IEEEauthorblockN{Rajesh Sundaresan}
\IEEEauthorblockA{Department of ECE\\
Indian Institute of Science\\
Bangalore, Karnataka 560012, India\\
Email: rajeshs@ece.iisc.ernet.in}
}

\maketitle

\begin{abstract}
This paper extends some geometric properties of a one-parameter family of relative entropies. These arise as redundancies when cumulants of compressed lengths are considered instead of expected compressed lengths. These parametric relative entropies are a generalization of the Kullback-Leibler divergence. They satisfy the Pythagorean property and behave like squared distances. This property, which was known for finite alphabet spaces, is now extended for general measure spaces. Existence of projections onto convex and certain closed sets is also established. Our results may have applications in the \renyi entropy maximization rule of statistical physics.
\end{abstract}

\section{Introduction}
\label{sec:introduction}

Relative entropy or Kullback-Leibler divergence
%\footnote{Relative entropy also called Kullback-Leibler (KL) information number, cross entropy, $I$-divergence, KL-divergence, or information divergence}
$I(P\|Q)$ between two probability measures is a fundamental quantity that arises in a variety of situations in probability, statistics, and information theory. It serves as a measure of dissimilarity or divergence between two probability measures $P$ and $Q$ on a given measure space. In information theory, it is well known that $I(P\|Q)$ is the penalty in expected compressed length, i.e., its gap from Shannon entropy $H(P)$, when the compressor assumes that the (finite-alphabet) source probability measure is $Q$ instead of the true probability measure $P$.

\renyi entropies $H_{\alpha}(P)$ for $\alpha \in (0, \infty)$ play the role of Shannon entropy when the normalized cumulant of compression length is considered instead of expected compression length. Indeed, Campbell \cite{1965xxIC_Cam} showed that
\[
  \min \frac{1}{n\rho} \log \mathbb{E} \left[ \exp \{ \rho L_n (X^n) \} \right] \to H_{\alpha} (P) ~ (\mbox{as } n \to \infty)
\]
for an independent and identically distributed (iid) source with marginal $P$. The minimum is over all compression strategies that satisfy the Kraft inequality, $\alpha = 1/(1+\rho)$, and $\rho > 0$ is the cumulant parameter. We also have $\lim_{\alpha \to 1} H_{\alpha}(P) = H(P)$, so that \renyi entropy may be viewed as a generalization of Shannon entropy.

If the compressor assumed that the true probability measure had marginal $Q$, instead of $P$, then the gap in the normalized cumulant's growth exponent from the optimal value (\renyi entropy) is an analogous parametric divergence quantity (introduced by Blumer and McEliece \cite{198809TIT_BluMcE} and studied further by Sundaresan \cite{200701TIT_Sun}), which we shall denote $I_{\alpha}(P,Q)$. The same quantity also arises when we study the gap from optimality of mismatched guessing exponents (see Arikan \cite{199601TIT_Ari} as well as Sundaresan \cite{200701TIT_Sun}). All these results are applicable to more general non-iid sources.

As one might expect, it is known that (see for example, Johnson and Vignat \cite[A.1]{200705AIHP_JohVig}) $\lim_{\alpha \to 1} I_{\alpha}(P,Q) = I(P\|Q)$, so that we may think of relative entropy as $I_1(P,Q)$, and therefore $I_{\alpha}$ as a generalization of relative entropy, i.e., an $\alpha$-relative entropy. Furthermore, for probability measures on a finite alphabet set, $I_{\alpha}$ behaves like squared Euclidean distance, and satisfies a ``Pythagorean property'' \cite{200701TIT_Sun} like relative entropy and squared Euclidean distance. One purpose of this paper is to extend this property to probability measures on a general measure space with some common dominating measure.

The maximum entropy principle is a well-known selection rule, in the presence of uncertainty, in statistics. For a source alphabet $\mathbb{X}$ with finite cardinality, by noting that $I(P \| U) = \log |\mathbb{X}| - H(P)$ with $U$ taken as the uniform measure on the finite alphabet set $\mathbb{X}$, the maximum entropy principle is the same as the minimum relative entropy principle, an idea that goes back to Boltzmann, and one which is supported by the theory of large deviations. Indeed, suppose that certain ensemble average measurements can be made on a realization of a sequence of iid random variables (mean, second moment, etc.). The resulting realization must have an empirical measure that obeys the constraints placed by the observations. In particular, the empirical measure belongs to a convex (and possibly closed) set. Large deviations theory tells us that, amongst the measures that respect the constraints, the one that minimizes relative entropy is exponentially more likely than the others. The resulting measure is called $I$-projection and was extensively studied by \csiszar \cite{1975xxAP_Csi}, \cite{1984xxAP_Csi}, and more recently by \csiszar and Mat\'{u}\v{s} \cite{200306TIT_CsiMat}. $I$-minimization arises similarly in the contraction principle of large deviations theory (see for example Dembo and Zeitouni's \cite[p.126]{1998xxLDTA_DemZei}).

As a natural alternative selection principle, the maximum \renyi entropy principle has been recently considered. This principle is equivalent to maximizing the Tsallis entropy, which is a monotone function of the \renyi entropy. See for example Jizba and Arimitsu \cite{2004xxJizAri}, and references therein. More interestingly, Jizba and Arimitsu \cite{2004xxJizAri} indicate that maximum \renyi entropy principle may be viewed as a maximum Shannon entropy principle with multifractal constraints. This selection principle has been of recent interest in statistical physics settings because \renyi entropy maximizers under a covariance constraint are distributions with a power-law decay (when $\alpha > 1$). See Costa et al. \cite{200307EMMCVPR_CosHerVig} or Johnson and Vignat \cite{200705AIHP_JohVig}. Several empirical observations in naturally arising physical and socio-economic systems possess a power-law decay. Without going into these aspects, we remark that $I_{\alpha} (P,U) = \log |\mathbb{X}| - H_{\alpha}(P)$, so that both the maximum \renyi entropy principle and the maximum Tsallis entropy principle are equivalent to a minimum ${\alpha}$-relative entropy (minimum $I_{\alpha}$) principle. Thus one needs to find amongst empirical measures that meet the observation constraints, the one that minimizes $I_{\alpha}$. We shall call this the $I_{\alpha}$-projection. While existence and uniqueness of $I_{\alpha}$-projection was proved by Sundaresan \cite{200701TIT_Sun} for the finite alphabet case, the second purpose of this paper is to extend these results to more general measure spaces.

It is known (see for example \cite{200701TIT_Sun}) that $I_{\alpha}(P,Q)$ is the more commonly studied \renyi divergence of order $1/\alpha$, not of the original measures $P$ and $Q$, but of their tilts $P'$ and $Q'$, where $P'(x) = P(x)^{\alpha} / Z(P)$, and $Z(P)$ is the normalization that makes $P'$ a probability measure. $Q'$ is similarly defined. While the \renyi divergences arise naturally in hypothesis testing problems (see for example \csiszar \cite{199501TIT_Csi}), $I_{\alpha}$ arises more naturally as a redundancy for mismatched compression.

$I_{\alpha}$ is also a certain monotone function of Csisz\'{a}r's $f$-divergence between $P'$ and $Q'$. As a consequence of the appearance of the tilts, the data-processing property satisfied by $f$-divergences does not hold for the $\alpha$-relative entropy. Surprisingly though, the Pythagorean property holds.

The rest of the paper is organized as follows. In section \ref{sec:projection}, we provide the definitions and demonstrate the existence of $I_{\alpha}$ projections on certain closed and convex sets. In section \ref{sec:pythagorenproperty}, we extend the Pythagorean property to general measure spaces (with a common dominating measure), and identify the consequences with respect to iterated projections. In section \ref{sec:concludingRemarks}, we summarize our results.

\section{$I_{\alpha}$-projection}
\label{sec:projection}

We first formalize the definition of $\alpha$-relative entropy to a general probability space.

Let $P$ and $Q$ be two probability measures on a measure space $(\mathbb{X},\mathcal{X})$. Let $\alpha \in (0,\infty)$ with $\alpha \neq 1$. By setting $\alpha=1/(1+\rho)$ we have the reparameterization in terms of $\rho$ with $-1<\rho<\infty$ and $\rho \neq 0$. Let $\mu$ be a dominating $\sigma$-finite measure on $(\mathbb{X},\mathcal{X})$ with respect to which $P$ and $Q$ are both absolutely continuous, denoted $P \ll \mu$ and $Q \ll \mu$. We denote $p = dP/d\mu$ and $q= dQ/d\mu$ and assume that they are in the complete metric space $L^{\alpha}(\mu)$ with metric
$$d(f,g) = \left( \int |f - g|^{\alpha} d\mu \right)^{\min\{1,1/\alpha \}}.$$
We shall use the notation
$$\| f \| := \left( \int |f|^{\alpha} d\mu \right)^{1/\alpha}$$
even though it is not a norm for $\alpha < 1$. (The dependence of this quantity on $\alpha$ should be borne in mind). The \renyi entropy of $P$ of order $\alpha$ (with respect to $\mu$) is given by
\[
  H_{\alpha}(P) = \frac{1}{1-\alpha} \log \left( \int p^\alpha d\mu \right).
\]
Consider the tilted measures $P'$ and $Q'$ given by
\begin{eqnarray*}
  \frac{dP'}{d\mu} = p' := \frac{p^{\alpha}}{\int p^{\alpha}d\mu} \mbox{ and } \frac{dQ'}{d\mu} = q' := \frac{q^{\alpha}}{\int q^{\alpha}d\mu}.
\end{eqnarray*}
$P'$ and $Q'$ are also dominated by $\mu$. With
$$f(x) := \text{sgn}(\rho) \cdot x^{1+\rho},$$
Csisz\'{a}r's $f$-divergence \cite{1967xxSSMH_Csi} between two measures $P$ and $Q$, both absolutely continuous with respect to $\mu$, is given by
\[
  I_{f}(P,Q) :=  \int q f \left( \frac{p}{q} \right) d\mu.
\]
Since $f$ is strictly convex when $\rho \neq 0$, by Jensen's inequality, $I_f(P,Q) \geq f(1)$ with equality if and only if $P=Q$.

We now define the $\alpha$-relative entropy to be
\[
  I_{\alpha}(P,Q) := \frac{1}{\rho} \log\left[ \text{sgn}(\rho) \cdot I_{f}(P',Q') \right].
\]
Abusing notation a little, when speaking of densities, we shall some times write $I_{\alpha}(p,q)$ for $I_{\alpha}(P,Q)$.

We now summarize the anticipated properties of $\alpha$-relative entropy.
\begin{lemma}
The following properties hold.

1) $I_{\alpha}(P,Q) \geq 0$ with equality if and only if $P = Q$.

2) Under certain regularity conditions, $\lim_{\alpha \to 1} I_{\alpha}(P,Q) = I(P \| Q)$.

3) Let $\mathbb{X} = \mathbb{R}^n$ and let $\mu$ be the Lebesgue measure on $\mathbb{R}^n$. For $\alpha > n / (n+2)$ and $\alpha \neq 1$, define the constant $b_{\alpha} = (1-\alpha)/(2\alpha - n(1-\alpha))$. With $C$ a positive definite covariance matrix, the function
      \[
        g_{\alpha, C}(x) = Z_{\alpha}^{-1} \left[ 1 + b_{\alpha} \cdot x^T C^{-1} x \right]^{\frac{1}{\alpha-1}}_+,
      \]
      with $[a]_+ := \max \{a,0 \}$ and $Z_{\alpha}$ the normalization constant, is the density function of a probability measure on $\mathbb{R}^n$ whose covariance matrix is $C$. Furthermore, if $g$ is the density function of any other random variable with covariance matrix $C$, then
      \begin{equation}
        \label{eqn:moment-entropy}
        I_{\alpha}(g, g_{\alpha,C}) = H_{\alpha}(g_{\alpha,C}) - H_{\alpha}(g).
      \end{equation}
      Consequently $g_{\alpha,C}$ is the density function of the \renyi entropy maximizer among all $\mathbb{R}^n$-valued random vectors with covariance matrix $C$.

4) Let $|\mathbb{X}| < \infty$ and let $U$ be the uniform probability mass function on $\mathbb{X}$. Then $I_{\alpha}(P,U) = \log |\mathbb{X}| - H_{\alpha}(P).$ $\hfill \IEEEQEDopen$
\end{lemma}

\begin{IEEEproof}
  We only give an outline here. Statement 1) follows by an application of H\"{o}lder's inequality by considering the H\"{o}lder conjugates $\alpha$ and $\alpha/(\alpha - 1)$, and the functions $p/\|p\|$ and $(q/\|q\|)^{\alpha-1}$. Statement 2) follows by an application of L'H\^{o}pital's rule and some conditions that enable interchange of differentiation with respect to the parameter $\alpha$ and integration with respect to $\mu$. Statement 3) was proved by Lutwak et al. \cite{200501TIT_LutYanZha}. See also Johnson and Vignat \cite{200705AIHP_JohVig}. For relative entropy, the analog of (\ref{eqn:moment-entropy}) under a covariance constraint would be $I(g \| \phi) = H(\phi) - H(g)$, where $H$ is differential entropy and $\phi$ is the Gaussian distribution with the same covariance as $g$. The last statement follows from the definition.
\end{IEEEproof}

We next prove an inequality relating $f$-divergences. This yields parallelogram identity for relative entropy ($\alpha = 1$) \cite{1975xxAP_Csi}.

\begin{lemma}
\label{parallel}
Let $\alpha < 1$. Let $P_1, P_2, R$ be probability measures that are absolutely continuous with $\mu$, and let the corresponding Radon-Nikodym derivatives $p_1, p_2,$ and $r$ be in $L^{\alpha}(\mu)$. Assume $0 \le \lambda \le 1$. We then have
\begin{align}
  \lefteqn{\lambda [I_f(P'_1,R')-f(1)]+(1-\lambda )[I_f(P'_2,R')-f(1)]} \nonumber \\
    & -\lambda  [I_f(P'_1,R'_{1,2})-f(1)]-(1-\lambda )[I_f(P'_2,R'_{1,2})-f(1)] \nonumber \\
    \label{eqn:parallelogram}
    & \ge  [I_f(R'_{1,2},R')-f(1)],
\end{align}
where
\begin{equation}
  \label{eqn:rstar}
  R_{1,2} = \displaystyle \frac{\lambda \frac{P_1}{\|p_1\|}+(1-\lambda)\frac{P_2}{\|p_2\|}}{\frac{\lambda}{\|p_1\|}+\frac{1-\lambda}{\|p_2\|}}.
\end{equation}
When $\alpha > 1$, the reversed inequality holds in (\ref{eqn:parallelogram}). $\hfill \IEEEQEDopen$
\end{lemma}
\begin{IEEEproof}
We briefly outline the steps. Let $r_{1,2} = dR_{1,2}/d\mu$. Observe that since $I_f(\cdot, \cdot) \geq f(1)$, a consequence of Jensen's inequality indicated earlier, all terms within square brackets are nonnegative. The left-hand side of inequality can be expanded to
\begin{eqnarray*}
  \lefteqn{ \text{sgn}(\rho) \int \frac{\lambda p_1}{\|p_1\|}
                                  \left[ \left( \frac{r}{\|r\|} \right)^{\alpha-1}
                                         - \left( \frac{r_{1,2}}{\| r_{1,2} \|} \right)^{\alpha-1}
                                  \right] d\mu } \\
  & & \hspace*{-.1in}+ \text{sgn}(\rho) \int \frac{(1-\lambda) p_2}{\|p_2\|}
                                  \left[ \left( \frac{r}{\|r\|}\right)^{\alpha-1}
                                         \hspace*{-.1in} - \left( \frac{r_{1,2}}{\| r_{1,2} \|} \right)^{\alpha-1}
                                  \right] d\mu \\
  & = & \text{sgn}(\rho) \int \frac{r_{1,2}}{\| r_{1,2} \|}
                                 \left[ \left( \frac{r}{\|r\|} \right)^{\alpha-1}
                                        - \left( \frac{r_{1,2}}{\| r_{1,2} \|} \right)^{\alpha-1}
                                 \right] d \mu\\
  & & \times \left[ \frac{\lambda}{\|p_1\|}+\frac{1-\lambda}{\|p_2\|} \right] \| r_{1,2}\| \\
  & = & \left[ \frac{\lambda}{\|p_1\|}+\frac{1-\lambda}{\|p_2\|} \right] \| r_{1,2}\| \cdot [ I_f(R'_{1,2}, R')-f(1) ].
\end{eqnarray*}
Applying Minkowski's inequality in (\ref{eqn:rstar}) with $\alpha < 1$, we get
\[
  \left(\frac{\lambda}{\|p_1\|}+\frac{1-\lambda}{\|p_2\|}\right) \|r_{1,2}\| \ge 1.
\]
This inequality gets reversed when $\alpha > 1$, again by a version of Minkowski's inequality. Since $I_f(R'_{1,2}, R')-f(1) \geq 0$, the lemma follows.
\end{IEEEproof}

Let us define what we mean by an $I_{\alpha}$-projection.
\begin{definition}
  If $E$ is a set of probability measures on $(\mathbb{X},\mathcal{X})$ such that $I_{\alpha}(P,R)<\infty$ for some $P\in E$, a measure $Q \in E$ satisfying
  \begin{eqnarray}
    \label{eqn:projection}I_{\alpha}(Q,R)=\displaystyle \inf_{P\in E}I_{\alpha}(P,R)
  \end{eqnarray}
  is called the $I_{\alpha}$-projection of $R$ on $E$. $\hfill \IEEEQEDopen$
\end{definition}

Let $E$ be a set of probability measures on $(\mathbb{X}, \mathcal{X})$. Let $\mu$ be a common ($\sigma$-finite) dominating measure for $E$. Write
\[
  \mathcal{E} = \left\{ p = \frac{dP}{d\mu} : P \in E \right\}
\]
and assume that $\mathcal{E} \subset L^{\alpha}(\mu)$. Now define
\[
  \mathcal{E'} := \left\{ p' = \left( \frac{p}{\|p\|} \right)^{\alpha} : p \in \mathcal{E} \right\}.
\]

We are now ready to state our main result on the existence of $I_{\alpha}$-projection.

\begin{thm}
\label{thm:min}
Let $\alpha \in (0, \infty)$ and $\alpha \neq 1$. Let $E$ be a set of probability measures with dominating $\sigma$-finite measure $\mu$ such that the subset of functions $\mathcal{E}$ is convex and closed in $L^{\alpha}(\mu)$. Let $R$ be a probability measure and suppose that $I_{\alpha}(P,R) < \infty$ for some $P \in E$. Then $R$ has an $I_{\alpha}$-projection on $E$. $\hfill \IEEEQEDopen$
\end{thm}

\begin{remark}
The closure of $\mathcal{E}$ in $L^{\alpha}(\mu)$, for $\alpha = 1$, would be closure in the total variation metric, which is one of the hypotheses in Csisz\'{a}r's \cite[Th.2.1]{1975xxAP_Csi}. The proof ideas are different for the two cases $\alpha < 1$ and $\alpha > 1$. The proof for $\alpha < 1$ is a modification of Csisz\'{a}r's approach in \cite{1975xxAP_Csi}. The proof for $\alpha > 1$ exploits properties of sets that are convex and closed under the weak topology. We are indebted to Pietro Majer for suggesting some key steps on the {\tt mathoverflow.net} forum.
\end{remark}

\begin{IEEEproof}
(a) We first consider the case $\alpha < 1$. Pick a sequence $P_n \in E$ such that $I_f(P'_n,R') < \infty$ and
\begin{eqnarray}
  \label{eqn:inf}
  I_f(P_n',R')\to\displaystyle\inf_{P\in E}I_f(P',R').
\end{eqnarray}
By Lemma (\ref{parallel}), we have
\begin{eqnarray}
  \lefteqn{\lambda I_f(P_m',R')+(1-\lambda)I_f(P_n',R')} \nonumber \\
     & - & \lambda I_f(P_m',R'_{m,n})-(1-\lambda) I_f(P_n',R'_{m,n}) \nonumber \\
     \label{eqn:projectionProofStep1}
     & & \ge ~ [I_f(R'_{m,n},R') - 1 ]
\end{eqnarray}
where
\[
  R_{m,n} =\displaystyle\frac{\lambda \frac{P_m}{\|p_m\|}+(1-\lambda)\frac{P_n}{\|p_n\|}}
                             {\frac{\lambda}{\|p_m\|}+\frac{1-\lambda}{\|p_n\|}} \in E
\]
on account of the convexity of $E$. Rearranging (\ref{eqn:projectionProofStep1}) and using $I_f(\cdot,\cdot) \geq f(1) = 1$, we get
\begin{eqnarray*}
  1  & \le & \lambda I_f(P'_m, R'_{m,n}) + (1-\lambda) I_f(P'_n,R'_{m,n}) \\
        & \le & \lambda I_f(P'_m, R') + (1-\lambda) I_f(P'_n, R') \\
        & & ~ - ~ [I_f(R'_{m,n}, R') - 1 ].
\end{eqnarray*}
Take the limit as $m,n \to \infty$. The expression on the right-most side is at most 1 because $I_f(P'_m, R')$ and $I_f(P'_n, R')$ approach the infimum value, and $I_f(R'_{m,n}, R')$ is at least this infimum value for each $m$ and $n$. Since we also have $I_f(P'_m, R'_{m,n}) \geq 1$ and $I_f(P'_n, R'_{m,n}) \geq 1$, it follows that
\[
  \lim_{m,n \to \infty} \left[ I_f(P_m',R'_{m,n}) - 1 \right] = 0.
\]
From \cite[Th. 1]{1967xxSSMH_Csi}, a generalization of Pinsker's inequality, we get that the total variation metric, denoted $|P - Q|$, is small if $I_f(P, Q) - 1$ is small. This fact and the above limit imply that
\[
  \lim_{m,n \to \infty} | P'_m - R'_{m,n} | = 0,
\]
which, together with the triangle inequality for the total variation metric, yields
\[
  |P_m'-P_n'| \le |P_n'-R'_{m,n}| + |P_m'-R'_{m,n}| \to 0 \mbox{ as } m,n \to \infty,
\]
i.e., the sequence $\{p_n'\}$ is a Cauchy sequence in $L^1(\mu)$. It must thus converge to some $g$ in $L^1(\mu)$, i.e.,
\begin{equation}
  \label{eqn:projectionProofStepk}
  \lim_{n \to \infty} \int \left| \left( \frac{p_n}{\|p_n\|} \right)^{\alpha} - g \right| d\mu = 0.
\end{equation}
There is then a subsequence, over which one gets a.e.$[\mu]$ convergence. Reindexing to operate on this subsequence, we get
\[
  \left( \frac{p_n}{\|p_n\|} \right)^{\alpha} \to g \mbox{ a.e.}[\mu].
\]
We will now demonstrate that an $I_{\alpha}$-projection, say $Q$, is in $E$ and has $\mu$-density proportional to $g^{1/\alpha}$.

In view of the a.e.$[\mu]$ convergence, and after observing that
\[
  \left| \frac{p_n}{\|p_n\|} - g^{1/\alpha} \right|^{\alpha} \le 2^{\alpha} \left[ \left( \frac{p_n}{\|p_n\|} \right)^{\alpha} + g \right],
\]
we can apply the generalized Dominated Convergence Theorem \cite[Ch.2, Problem.20]{1999xxRA_Fol} to get
\[
  g_n^{1/\alpha} := \frac{p_n}{\|p_n\|} \to g^{1/\alpha} \mbox{ in } L^{\alpha}(\mu).
\]

We next claim that
\begin{equation}
  \label{eqn:norm-pn-bounded}
  \|p_n\| \mbox{ is bounded.}
\end{equation}
Suppose not; then working on a subsequence if needed, we have $\|p_n\| := M_n \to \infty$. As $\int p_n d\mu=1$, given any $\epsilon>0$,
\[
  \mu(g_n>\epsilon^{\alpha})=\mu(p_n>\epsilon M_n)\le \frac{1}{\epsilon M_n}\to 0 \mbox{ as } n\to \infty,
\]
and hence $g_n\to 0$ in $[\mu]$-measure, which would be a contradiction to the fact that $\int g_n d\mu=1$ for all $n$. Thus (\ref{eqn:norm-pn-bounded}) holds, and so we can find a subsequence that converges to some $c$. Reindex and work on this subsequence to get $p_n \to cg^{1/\alpha}$ in $L^{\alpha}(\mu)$. Since $\mathcal{E}$ is closed in $L^{\alpha}(\mu)$, we obtain $cg^{1/\alpha}=q$ for some $q \in \mathcal{E}$, $c=\|q\|$, and $g=q^{\alpha}/\|q\|^{\alpha} \in \mathcal{E}'$. Let $Q$ be the probability measure in $E$ with $dQ/d\mu = q$.

To complete the proof, we need to demonstrate that $I_{\alpha}(P', R') \geq I_{\alpha}(Q', R')$ for every $P \in E$. To see this, note that (\ref{eqn:projectionProofStepk}) implies that $p'_n \to q'$ in $L^{1}(\mu)$, and by a change of measure, $p'_n/r' \to q'/r'$ in $L^{1}(R')$, and hence in $[R']$-measure. But $f$ is continuous, and so $f\left(p_n' / r' \right) \to f\left(q' / r' \right)$ in $[R']$-measure. Fatou's lemma then implies
\begin{eqnarray}
  \label{eqn:liminf}
  I_f(Q',R') \le \liminf_{n \to \infty} I_f(P_n', R') = \inf_{P \in E} I_f(P', R').
\end{eqnarray}
Since $Q\in E$, equality must hold, and $Q$ is an $I_{\alpha}$-projection of $R$ on $E$. This completes the proof for the case when $\alpha < 1$.

\vspace*{.1in}

(b) We next consider the case when $\alpha > 1$. Note that $\rho$ is negative, and so the $\inf$ in (\ref{eqn:projection}) becomes a $\sup$ as follows. The $I_{\alpha}$-projection $Q$ must satisfy (\ref{eqn:projection}) which can be rewritten as
\begin{eqnarray}
  \label{eqn:inftosup}
  I_{\alpha}(Q,R)
    & = & \frac{1}{\rho} \log \left[ \sup_{p \in \mathcal{E}} \int \frac{p}{\|p\|} \left( \frac{r}{\|r\|} \right)^{\alpha-1} d\mu \right] \\
    & = & \frac{1}{\rho} \log \left[ \sup_{h \in \hat{\mathcal{E}}} \int h g ~ d\mu \right], \nonumber
\end{eqnarray}                                                                                                                                                                                                                                                                                                                                                                                                       where
$$\hat{\mathcal{E}} := \left\{ s \frac{p}{\|p\|} : p \in \mathcal{E}, 0 \le s\le 1 \right\},$$
and $g=\left(r / \|r\| \right)^{\alpha-1}$, an element of the dual space $\left(L^{\alpha}(\mu)\right)^*$.

We now claim that
\begin{equation}
  \label{eqn:claim-closed-convex}
  \hat{\mathcal{E}} \mbox{ is a closed and convex subset of } L^{\alpha}(\mu).
\end{equation}
Assume the claim. Since $L^{\alpha}(\mu)$ is a reflexive space, the closed and convex set $\hat{\mathcal{E}}$ is closed under the weak topology. Since $\hat{\mathcal{E}}$ is also contained in the unit sphere in $L^{\alpha}(\mu)$, the unit sphere being compact in the weak topology in a reflexive space, $\hat{\mathcal{E}}$ must be compact in the weak topology. The supremum is thus of a bounded linear functional over the weakly compact set $\hat{\mathcal{E}}$. It is therefore attained in $\hat{\mathcal{E}}$. Since the linear functional increases with $s$, the supremum is attained with $s=1$. Thus the supremum in (\ref{eqn:inftosup}) over $p \in \mathcal{E}$ is attained.

We now proceed to show the claim (\ref{eqn:claim-closed-convex}). To see convexity, let $p_1, p_2 \in \mathcal{E}$ and $0 < s_1, s_2, \lambda <1$. Then
\begin{eqnarray*}
  \lefteqn { \lambda s_1 \frac{p_1}{\|p_1\|} + (1-\lambda) s_2 \frac{p_2}{\|p_2\|} } \\
  & & = \left( \frac{\lambda s_1}{\|p_1\|} + \frac{(1-\lambda)s_2}{\|p_2\|} \right) \cdot \|p\| \cdot \frac{p}{\|p\|} \\
  & & =: s \frac{p}{\|p\|}
\end{eqnarray*}
where
\[
  p := \frac{\frac{\lambda s_1}{\|p_1\|}p_1 + \frac{(1-\lambda)s_2}{\|p_2\|}p_2}
            {\frac{\lambda s_1}{\|p_1\|} + \frac{(1-\lambda)s_2}{\|p_2\|}}
    \in \mathcal{E}
\]
by the convexity of $\mathcal{E}$. From Minkowski's inequality (for $\alpha > 1$), we also have
\[
  s = \left( \frac{\lambda s_1}{\|p_1\|} + \frac{(1-\lambda)s_2}{\|p_2\|} \right) \cdot \|p\| \le \lambda s_1 + (1- \lambda) s_2 \leq 1,
\]
and this establishes the convexity of $\hat{\mathcal{E}}$.

To see that $\hat{\mathcal{E}}$ is closed in $L^{\alpha}(\mu)$, let $\{ g_n \} \subset \hat{\mathcal{E}}$ be a Cauchy sequence in $L^{\alpha}(\mu)$. Then $g_n = s_n p_n / \|p_n\|$, with $p_n \in \mathcal{E}$ and $0 \leq s_n \leq 1$, converges to some $g$ in $L^{\alpha}(\mu)$. By taking norms, we see that $\| g_n \| = s_n \to \| g \| \leq 1$. If $g=0$ a.e.$[\mu]$, then $g \in \hat{\mathcal{E}}$ by taking $s = 0$, and we are done. Otherwise we can assume that $\| g_n \| > 0$ for all $n$ by focusing on a subsequence if needed, and that $\|g\| > 0$. We can thus conclude that $p_n / \|p_n\| = g_n / \|g_n\| \to g / \|g\|$ in $L^{\alpha}(\mu)$. Since $g \neq 0$, the same argument that showed (\ref{eqn:norm-pn-bounded}) shows that $\|p_n\|$ is bounded, and by focusing on a subsequence, we may assume that it converges to some constant $c$. Hence $p_n \to cg/\|g\|$ in $L^{\alpha}(\mu)$. Since $\mathcal{E}$ is closed, we must have $cg/\|g\|=p$ for some $p\in \mathcal{E}$, $c=\|p\|$, and $g=\|g\| p / \|p\|$. Since we already established that $\|g\| \leq 1$, it follows that $g \in \hat{\mathcal{E}}$. This completes the proof.
\end{IEEEproof}

We close this section with a result on the continuity or the lower semicontinuity of $\alpha$-relative entropy.

\begin{proposition}
For a fixed $q$, consider $p \mapsto I_{\alpha}(p,q)$ as a function on $L^{\alpha}(\mu)$. This function is continuous for $\alpha>1$ and lower semicontinuous for $\alpha<1$. $\hfill \IEEEQEDopen$
\end{proposition}

\begin{IEEEproof}
Let us first consider the case when $\alpha>1$. Let $p_n\to p$ in  $L^{\alpha}(\mu)$. Then $\|p_n\|\to \|p\|$ and so $p_n/\|p_n\| \to p/\|p\|$ in  $L^{\alpha}(\mu)$. As mentioned in the proof of Theorem \ref{thm:min}(b),  $I_{\alpha}(p,q)$ is a monotone function of a bounded linear functional in $p/\|p\|$. Hence $ I_{\alpha}(p,q)$ is continuous in $p$. For $\alpha<1(\rho>0)$ we write
\[
  I_{\alpha}(p,q)=\frac{1}{\rho}\log \left[ \int (p'/q')^{1+\rho} dQ' \right].
\]
Let $p_n\to p$ in $L^{\alpha}(\mu)$. Then $\|p_n\| \to \|p\|$ and since $|p_n^{\alpha}-p^{\alpha}| \le |p_n|^{\alpha}+|p|^{\alpha}$, the generalized Dominated Convergence Theorem yields 
\[
 (p_n/\|p_n\|)^{\alpha} \to (p/\|p\|)^{\alpha} \mbox{ in } L^1(\mu),
\]
i.e., $p_n' \to p'$ in $L^1(\mu)$. This is the same as saying $p_n' / q' \to p' / q'$ in $L^1(Q')$, and thus in $[Q']$-measure. Hence it follows that $(p_n' / q')^{1+\rho} \to (p' / q')^{1+\rho}$ in $[Q']$-measure. By Fatou's lemma, 
\[
\liminf_{n\to\infty} \int (p_n'/q')^{1+\rho}dQ'\ge \int (p'/q')^{1+\rho}dQ'.
\]
As increasing function of a lower semicontinuous function is lower semicontinuous, the result is established for $\alpha<1$.
\end{IEEEproof}

\section{Pythagorean property}
\label{sec:pythagorenproperty}

In this section, we state the Pythagorean property for $\alpha$-relative entropy. We define the $I_{\alpha}$-sphere with center $R$ and radius $r$ as $S(R,r)=\{P:I_{\alpha}(P,R)<r\},~ 0<r\le \infty$. 

\begin{thm}
\label{thm:pythagorean}
Let $\alpha>0$ and $\alpha\neq 1$. Let $\mu$ be a common dominating $\sigma$-finite measure.
\begin{enumerate}
\item \label{item:inequality}If $I_{\alpha}(P,R)$ and $I_{\alpha}(Q,R)$ are finite, ``the segment joining $P$ and $Q$'' does not intersect the $I_{\alpha}$-sphere $B(R,r)$ with radius $r=I_{\alpha}(Q,R)$, i.e., $I_{\alpha}(P_{\lambda},R)\ge I_{\alpha}(Q,R)$ for
    \[
      P_{\lambda}=\lambda P+(1-\lambda)Q,~\lambda\in [0,1]
    \] 
    if and only if 
    \begin{eqnarray}
      \label{eqn:pythagoreaninequality}
      I_{\alpha}(P,R)\ge I_{\alpha}(P,Q)+I_{\alpha}(Q,R).
    \end{eqnarray}
\item \label{item:equality} If
    \begin{eqnarray}
      \label{eqn:convex combination} Q=\lambda P+(1-\lambda)S,~~ 0<\lambda<1 
    \end{eqnarray} 
    and 
    $I_{\alpha}(Q,R)$ is finite, then the segment joining $P$ and $S$ does not intersect $B(R,r)$ with $r=I_{\alpha}(Q,R)$, if and only if $I_{\alpha}(P,R)=I_{\alpha}(P,Q)+I_{\alpha}(Q,R)$ and $I_{\alpha}(S,R)= I_{\alpha}(S,Q)+I_{\alpha}(Q,R)$. $\hfill \IEEEQEDopen$
\end{enumerate}
\end{thm}

For the proof(see Appendix), we proceed as in \cite{200701TIT_Sun} where it is proved for the finite alphabet case, with appropriate functional analytic justifications for the general alphabet case.

Once Theorem \ref{thm:pythagorean} is established in generality, the proofs of the following results are exactly as in \cite{200701TIT_Sun}.

\begin{thm} The following statements hold.

1) ({\em Projection}) A $Q\in E \cap S(R,\infty)$ is an $I_{\alpha}$-projection of $R$ on the convex set $E$ iff every $P\in E$ satisfies (\ref{eqn:pythagoreaninequality}). If the $I_{\alpha}$-projection is an algebraic inner point of $E$ then $E \subset S(R,\infty)$ and (\ref{eqn:pythagoreaninequality}) holds with equality.

2) ({\em Uniqueness of $I_{\alpha}$-projection}) If $I_{\alpha}$-projection exists, it is unique.

3) ({\em Iterative projection}) Let $E$ and $E_1 \subset E$ be convex sets of probability measures, let $R$ have $I_{\alpha}$-projection $Q$ on $E$ and $Q_1$ on $E_1$, and suppose that (\ref{eqn:pythagoreaninequality}) holds with equality for every $P \in E$. Then $Q_1$ is the $I_{\alpha}$-projection of $Q$ on $E_1$. $\hfill \IEEEQEDopen$
\end{thm}

\section{Summary}
\label{sec:concludingRemarks}
We studied a parametric extension of relative entropy $I_{\alpha}$ for $\alpha > 0$ and $\alpha \neq 1$. These arose naturally as redundancies under mismatched compression and when normalized cumulants of compression lengths are considered ($0 \leq \alpha \leq 1$). We first studied $I_{\alpha}$ minimization problems and showed that projections exist on convex and closed sets (in $L^{\alpha}(\mu))$ when the sets are dominated by a $\sigma$-finite measure $\mu$. We then extended the Pythagorean property to general measure spaces. As a consequence, one also gets an iterated projections property. Axiomatic characterizations that lead to $I_{\alpha}$ minimization and \renyi entropy maximization are currently under investigation.

\section*{Acknowledgements}
The first author was supported by a Council for Scientific and Industrial Research (CSIR) fellowship. The work was supported in part by the University Grants Commission by Grant Part (2B) UGC-CAS-(Ph.IV).

\bibliographystyle{IEEEtran}
{
\bibliography{IEEEabrv,arxiv_final}
}

\newpage
\appendices

\section{Proof of Theorem \ref{thm:pythagorean}}

\ref{item:inequality}) We first prove statement \ref{item:inequality}). We begin with the ``only if'' part. Under the hypothesis, it suffices to show that
\[
  \text{sgn}(\rho)\cdot I_{f}(P',R')\ge \text{sgn}(\rho)\cdot I_{f}(P',Q')\cdot I_{f}(Q',R').
\]
Now,
\begin{eqnarray*}
  I_{f}(P',R')
    & = & \int r'f\left(\frac{p'}{r'}\right)d\mu\\
    & = & \text{sgn}(\rho)\cdot \int (p')^{1+\rho}(r')^{-\rho}d\mu\\&=& \frac{\text{sgn}(\rho)}{\|p\|} \cdot \int p~(r')^{-\rho} d\mu.
\end{eqnarray*}
Therefore it suffices to show that
\begin{eqnarray}
  \lefteqn{ \text{sgn}(\rho)\int p~(r')^{-\rho} d\mu} \nonumber \\
    \label{eqn:equivalentpythagoreanidentity}
    & \ge & \frac{\text{sgn}(\rho)}{\|q\|}\int p~ (q')^{-\rho} d\mu
    \cdot \int q~(r')^{-\rho} d\mu
\end{eqnarray}
Now
\begin{eqnarray*}
  I_{f}(P_{\lambda}',R')
    & = & \frac{\text{sgn}(\rho)}{\|p_{\lambda}\|}\cdot \int p_{\lambda}\cdot (r')^{-\rho} d\mu \\
    & =: & \frac{s(\lambda)}{t(\lambda)}
\end{eqnarray*}
where
\[
  s(\lambda) := \text{sgn}(\rho)~\int p_{\lambda}\cdot (r')^{-\rho}d\mu \mbox{ and } t(\lambda) := \|p_{\lambda}\|.
\]
Clearly, $I_{\alpha}(P_{\lambda},R)\ge I_{\alpha}(Q,R)$ for $\lambda \in (0,1)$ implies that
\begin{eqnarray} \label{eqn:derivativeoff-divergence}\frac{I_{f}(P_{\lambda}',R')-I_{f}(P_{0}',R')}{\lambda}\ge 0 ~~\text{for}~~ \lambda \in (0,1).\end{eqnarray}
Therefore the limiting value as $\lambda \downarrow 0$, the derivative of $I_f(P_{\lambda}',R')$ with respect to $\lambda$ evaluated at $\lambda=0$, should be $\ge 0$. We then have
\begin{eqnarray*}
  \frac{s(\lambda)-s(0)}{\lambda}
   & = & \frac{\text{sgn}(\rho)}{\lambda}\left[\int p_{\lambda}(r')^{-\rho}d\mu-\int q~(r')^{-\rho}d\mu\right]\\
   & = & \text{sgn}(\rho)~\int \left(\frac{p_{\lambda}-q}{\lambda}\right)~ (r')^{-\rho}d\mu\\
   & = & \text{sgn}(\rho)~\int (p-q)~(r')^{-\rho}d\mu\\&=&\text{sgn}(\rho)\left[\int p~(r')^{-\rho}d\mu-\int q~(r')^{-\rho}d\mu\right].
\end{eqnarray*}
So $\dot{s}(0) = \lim_{\lambda \downarrow 0} (s(\lambda)-s(0))/\lambda$ exists and equals the above expression. For $\alpha>1$, we have
\[
  \left| \frac{\partial{}}{\partial \lambda}(p_{\lambda})^{\alpha} \right| = \alpha |p-q|(p_\lambda)^{\alpha-1}\le \alpha (p+q)^{\alpha},
\]
while for $\alpha<1$, we have
\begin{eqnarray*}
  \left| \frac{\partial{}}{\partial \lambda}(p_{\lambda})^{\alpha} \right| 
    =  \alpha~ |p-q|(p_\lambda)^{\alpha-1} 
    \le \frac{\alpha~(p+q)^{\alpha}}{\{\min{(\lambda, 1-\lambda)}\}^{1-\alpha}},
\end{eqnarray*}
and both upper bounds are in $L^1(\mu)$ for a fixed $\lambda > 0$. Therefore by chain rule and \cite[Th. 2.27]{1999xxRA_Fol}, we get
\[
  \dot{t}(\lambda) = \left[ \int (p_{\lambda})^{\alpha} d\mu \right]^{\frac{1}{\alpha}-1} \cdot \int (p_{\lambda})^{\alpha-1}(p-q) d\mu
\]
for each $\lambda > 0$. Taking $\lambda \downarrow 0$, we get
\begin{eqnarray*}
  \dot{t}(0)
    & = & \left(\int q^{\alpha}d\mu\right)^{\frac{1}{\alpha}-1}\cdot\int q^{\alpha-1}(p-q)d\mu \\
    & = & \left(\int q^{\alpha}d\mu\right)^{\frac{1-\alpha}{\alpha}}\cdot\left(\int pq^{\alpha-1}d\mu-\int q^{\alpha}d\mu\right) \\
    & = & \int p\left(\frac{q^{\alpha}}{\int q^{\alpha}d\mu}\right)^{\frac{\alpha-1}{\alpha}}d\mu-\left(\int q^{\alpha}d\mu\right)^{\frac{1}{\alpha}} \\
    & = & \int p\cdot (q')^{-\rho}d\mu - \|q\|.
\end{eqnarray*}
Thus
\begin{eqnarray*}
  \lefteqn{\displaystyle\frac{1}{\lambda}\left[\frac{s(\lambda)}{t(\lambda)}-\frac{s(0)}{t(0)}\right]} \\
  & = & \frac{1}{t(\lambda)t(0)} \left[t(0)\frac{s(\lambda)-s(0)}{\lambda}-s(0)\frac{t(\lambda)-t(0)}{\lambda}\right].
\end{eqnarray*}
It follows that the derivative of $s(\lambda)/t(\lambda)$ exists at $\lambda=0$ and is given by $(t(0)\dot{s}(0)-s(0)\dot{t}(0))/t^2(0)$. Equation (\ref{eqn:derivativeoff-divergence}) together with  $t(0)>0$ imply that
\begin{eqnarray}
  \label{eqn:nonnegativity}
  \dot{s}(0)-s(0)\cdot\frac{\dot{t}(0)}{t(0)}\ge0.
\end{eqnarray}
Consequently, $\dot{t}(0)$ is necessarily finite. Substituting the values of $s(0), \dot{s}(0), t(0)$ and $\dot{t}(0)$ in (\ref{eqn:nonnegativity}) we get the required inequality (\ref{eqn:equivalentpythagoreanidentity}).

To prove the converse ``if'' part, let us assume that 
\[
  I_{\alpha}(P,R)\ge I_{\alpha}(P,Q)+I_{\alpha}(Q,R),
\]
which is the same as (\ref{eqn:equivalentpythagoreanidentity}). It also implies that $I_{\alpha}(P,Q)$ is also finite. From the trivial statement $I_{\alpha}(Q,R)=I_{\alpha}(Q,Q)+I_{\alpha}(Q,R)$, we have
\begin{eqnarray}
  \lefteqn{ \text{sgn}(\rho) \int q~(r')^{-\rho} d\mu } \nonumber \\
  \label{eqn:equality}
  & = & \frac{\text{sgn}(\rho)}{\|q\|} \cdot \int q~ (q')^{-\rho} d\mu \cdot \int q~(r')^{-\rho} d\mu.
\end{eqnarray}
A $\lambda$-weighted linear combination of (\ref{eqn:equivalentpythagoreanidentity}) and (\ref{eqn:equality}) yields,
\begin{eqnarray*}
  \lefteqn{ \text{sgn}(\rho) \int p_{\lambda}~(r')^{-\rho} d\mu } \\
    & \ge & \frac{\text{sgn}(\rho)}{\|q\|} \cdot \int p_{\lambda}~(q')^{-\rho} d\mu \cdot \int q~(r')^{-\rho} d\mu,
\end{eqnarray*}
i.e.,
\begin{eqnarray*}
  I_{\alpha}(P_{\lambda},R)&\ge& I_{\alpha}(P_{\lambda},Q)+I_{\alpha}(Q,R)\\&\ge&I_{\alpha}(Q,R).
\end{eqnarray*}

\ref{item:equality}) We next prove statement \ref{item:equality}). From $I_{\alpha}(Q,R)$ being finite, we claim that $I_{\alpha}(P,R)$ and $I_{\alpha}(S,R)$ are also finite. From (\ref{eqn:convex combination}), it is clear that $p/q\le {\lambda}^{-1}$ and thus $p/r \le {\lambda}^{-1} q/r$.
As a consequence, we have
\begin{eqnarray*} 
  \left(\frac{p'}{r'}\right)^{\frac{1}{\alpha}}
    & = & \frac{p}{r}\cdot \frac{\|r\|}{\|p\|} \\
    & \le & {\lambda}^{-1}\frac{q}{r}\cdot \frac{\|r\|}{\|p\|} \\
    & = & {\lambda}^{-1}\left(\frac{q'}{r'}\right)^{\frac{1}{\alpha}}\cdot \frac{\|q\|}{\|p\|}.
\end{eqnarray*}
Integrating with respect to $R'$, we get
\[
  \int \left(\frac{p'}{r'}\right)^{\frac{1}{\alpha}} dR' 
  \le {\lambda}^{-1}\frac{\|q\|}{\|p\|}\cdot \int \left(\frac{q'}{r'}\right)^{\frac{1}{\alpha}}dR'
  < \infty.
\]

\newpage
Taking the sign of $\rho$ appropriately, it immediately follows that $I_{\alpha}(P,R) \leq I_{\alpha}(Q,R) + $ finite constant, and is therefore finite. Similarly $I_{\alpha}(S,R)$ is also finite. Applying the first part of the theorem, we get
\begin{eqnarray*}
  I_{\alpha}(P,R) & \ge & I_{\alpha}(P,Q)+I_{\alpha}(Q,R) \\
  I_{\alpha}(S,R) & \ge & I_{\alpha}(S,Q)+I_{\alpha}(Q,R).
\end{eqnarray*}
If either of these were a strict inequality, then the linear combination $Q=\lambda P+(1-\lambda)S$ will satisfy (\ref{eqn:equality}) with strict inequality, a contradiction. So both the above must be equalities proving the ``only if'' part. The converse ``if'' part trivially follows from (\ref{item:inequality}). $\hfill \IEEEQEDclosed$

\end{document}